\documentclass[useAMS, usenatbib, usegraphicx]{mn2e}
\usepackage{amssymb}
\usepackage{booktabs}
\newcommand\ApJ{ApJ}

\def\alf{Alfv\'en\,}
\def\bq{\begin{equation}}
\def\eq{\end{equation}}
\def\ee #1 {\times 10^{#1}}
\def\ut #1 #2 { \, \rmn{#1}^{#2}}
\def\u #1 { \, \rmn{#1}}

\def\half{{\textstyle \frac{1}{2}}}

\let\grad=\nabla
\newcommand\cross{\bmath{\times}}

\newcommand\bm{\bmath}
\def\curl{{\grad \cross}}
\def\div #1 {\grad \cdot #1}

\def\cs{\cos\theta}

\def\b{\bmath{b}}

\def\v{\bmath{v}}

\def\k{\bmath{k}}
\def\kh{\bmath{\hat{k}}}

\def\x{\bmath{x}}
\def\vD{\bmath{v}_D}
\def\vi{\bmath{v}_i}
\def\vj{\bmath{v}_j}
\def\ve{\bmath{v}_e}
\def\vi{\bmath{v}_i}

\def\vn{\bmath{v}_n}

\def\B{\bmath{B}}
\def\hB{\hat{\B}}

\def\Bh{\widehat\B}
\def\E{\bmath{E}}            
\def\Bh{\bmath{\hat{B}}}

\def\J{\bmath{J}}

\def\drho{\delta\rho}
\def\dv{\bmath{\delta\v}}

\def\dB{\bmath{\delta\B}}
\def\dJ{\bmath{\delta\J}}

\def\sigv{<\sigma v>}

\newcommand{\delt} [1] {\frac{\partial #1}{\partial t}}

\def\b1{{\bar{\omega}}}
\def\bo2{\bar{\omega}^2}

\title{Hall magnetohydrodynamics of partially ionized plasmas}
\author[B.P.Pandey and Mark Wardle]
        {B.P. Pandey and Mark Wardle \\
{Department of Physics, Macquarie University, Sydney, NSW 2109, Australia} }
\date{\today}
\pagerange{\pageref{firstpage}--\pageref{lastpage}}
\pubyear{2007}
\begin{document}
\maketitle
\label{firstpage}
\begin{abstract}
The Hall effect arises in a plasma when electrons are able to drift with
the magnetic field but ions cannot.  In a fully-ionized plasma this
occurs for frequencies between the ion and electron cyclotron
frequencies because of the larger ion inertia.  Typically this
frequency range lies well above the frequencies of interest (such as
the dynamical frequency of the system under consideration) and can be
ignored.  In a weakly-ionized medium, however, the Hall effect arises
through a different mechanism -- neutral collisions preferentially
decouple ions from the magnetic field.  This typically occurs at much
lower frequencies and the Hall effect may play an important role in
the dynamics of weakly-ionised systems such as the Earth's ionosphere
and protoplanetary discs.

To clarify the relationship between these mechanisms we develop an
approximate single-fluid description of a partially ionized plasma
that becomes exact in the fully-ionized and weakly-ionized limits.
Our treatment includes the effects of ohmic, ambipolar, and Hall
diffusion.  We show that the Hall effect is relevant to the dynamics
of a partially ionized medium when the dynamical frequency exceeds the
ratio of ion to bulk mass density times the ion-cyclotron frequency,
i.e.\ the Hall frequency.  The corresponding length scale is inversely
proportional to the ion to bulk mass density ratio as well as to the
ion-Hall beta parameter.  In a weakly ionized medium, the critical
frequency becomes small enough that Hall MHD is an accurate
representation of the dynamics.  More generally, ohmic and ambipolar
diffusion may also be important.

We show that both ambipolar and Hall diffusion depend upon the
fractional ionization of the medium.  However, unlike ambipolar
diffusion, Hall diffusion may also be important in the high fractional
ionization limit.  The wave properties of a partially-ionized medium
are investigated in the ambipolar and Hall limits.  We show that in
the ambipolar regime wave damping is dependent on both fractional
ionization and ion-neutral collision frequencies.  In the Hall regime,
since the frequency of a whistler wave is inversely proportional to
the fractional ionization, and bounded by the ion-neutral collision
frequency it will play an important role in the Earth's ionosphere,
solar photosphere and astrophysical discs.

\end{abstract}


\section{Introduction}

In ideal magnetohydrodynamics (MHD), ions and electrons are both tied
to the magnetic field.  When ions are decoupled from the field and
electrons are not, the magnetic field and electrons drift together
through the ions and the generalised Ohm's law is modified by the Hall
electric field, proportional to $\J\cross\B$, where $\J$ and $\B$ are
the current density and magnetic field.  This modification of ideal
MHD is called Hall magnetohydrodynamics.

Hall MHD plays a crucial role in a variety of astrophysical, space and
laboratory environments, often providing the dominant mechanism for
plasma drift against the magnetic field, from flux expulsion in
neutron star crusts \citep{gr} to angular momentum transport in weakly
ionized protoplanetary discs (Wardle, 1999; Balbus and Terquem, 2001; hereafter W99 and BT01 respectively).  The formation of
intensive flux tubes in the solar atmosphere \citep{khod}, waves in
the solar wind \citep{zhel, mit}, propagation of whistlers in the
Earth's ionosphere \citep{abu} and sub-Alfv\'enic plasma expansion
\citep{huba} are but a few examples where Hall MHD appears to play
significant role. In fusion plasmas, the Hall effect can play an important role in describing various discharge behaviour \citep{KGK81, WB93}. For example, it can significantly enhance the non-Ohmic current drive in tokamaks \citep{pan}.

Two mechanisms may decouple the ions from the magnetic field under different physical conditions. This has led to distinct approaches being adopted to investigate the role of the Hall effect in the dynamics of laboratory \citep{ KGK81, WB93, pan}, space \citep{huba, hub2, abu, rich, zhel} and astrophysical (W99, BT01, Goldreich and Reisenegger, 1992) plasmas.

In a highly ionized plasma the Hall effect arises because of the
difference in electron and ion inertia: ions are unable to follow
magnetic fluctuations at frequencies higher than their cyclotron
frequency, whereas electrons remain coupled to the magnetic
field. The corresponding physical scale, the ion skin depth, is
typically much smaller than the scale of the system.  In this case
the Hall effect has typically been incorporated by explicitly
including the ion-electron drift in the induction equation.

In a partially ionized plasma the Hall effect may instead arise
because neutral collisions more easily decouple ions from the
magnetic field than electrons. In this case, the Hall scale can
become comparable to the size of the system itself. Its effects are
typically incorporated through a second-rank conductivity tensor
appearing in a generalized Ohm's law \citep{cow, mich}.

The Hall dynamics of highly ionized and weakly ionized plasmas
are similar, but occur on very different frequency ranges and
spatial scales due to the different mechanisms responsible for the
underlying symmetry breaking in ion and electron dynamics.  This has
led to some confusion in the literature, where estimates of the
fully-ionised Hall length scale have been applied to the ionized
component of partially ionized media to conclude that the Hall
effect is irrelevant in circumstances when it is, in fact, crucial
\citep{huba, hub2, BCP97, rud01}. 

The purpose of this paper is to clarify the relationship between the
fully ionized and weakly ionised limits by developing a unified
single-fluid framework for the dynamics of plasmas of arbitrary
ionisation.  Our treatment is of necessity approximate in the
intermediate case, but has the correct behaviour in the highly- or
weakly-ionised limits and is not strongly limited in applicability
in the intermediate ionisation regime.  This allows us to explore
the change of scale in the Hall effect in moving from fully to
partially ionized plasmas and gain a deep physical understanding of
the nature of the transition between the two ionisation regimes.
Furthermore, this formulation is useful in gaining insight into the
behaviour of plasmas that are neither fully ionized nor weakly
ionized (e.g. near a tokamak wall or the surface of a white dwarf),
when neutral collisions and ionized plasma inertia may both be
important.  

The paper is organised in the following fashion. In section 2 we
derive a set of fluid equation in the bulk frame suitable for the
weakly ionized medium and the
characteristic scales on which the Hall effect manifests, are
discussed.  In section 3, waves in a partially ionized plasma are
described and the dependence of the wave damping on fractional
ionization in the ambipolar regime is discussed. The very low frequency 
ion-cyclotron and high frequency collisional whistler is shown to be the two branches in the Hall regime.  In
section 4 we discuss the potential wide ranging applications of this
work to laboratory, space, and astrophysical plasmas.  A brief
summary of the results is given in the final section.

\section{Formulation}
Space and astrophysical plasmas are generally partially ionized
consisting of electrons, ions, neutrals, and charged and neutral
dust grains.  We shall neglect grains in the present formulation and
consider a partially ionized plasma consisting of electrons, ions,
and neutrals.  The dynamics of such a plasma is complex but depending 
upon the physical conditions pertaining to the problem at hand, reasonable
simplifying assumptions can be made.  For example, the dynamics of a
protoplanetary disc has been investigated by assuming that the
neutrals provide the inertia of the bulk fluid and plasma particles
carry the current (W99).  This approach is reasonable as in a
cold protostellar disc, the ionization fraction (i.e. the ratio of
electron to the neutral number density) is very low ($\sim
10^{-8}-10^{-13}$) and the relative drift between ions and neutrals are
small.  Therefore, such a description is not only economical but
also captures the essential physics of the protoplanetary discs.
However, the inertia of the ionized components may in general play
an important role, e.g. near the wall of a tokamak, in the lower
part of Earth's F-region, at the base of the solar chromosphere, in
the outer part of AGN discs, in the discs around the dwarf novae
etc., when the ionization fraction is small and yet not negligible.
In neutron star crusts too, neutron and proton densities are
comparable and a multi-component description of the strongly
magnetized fluid is desirable.  In the solar chromosphere, utilizing
three component description, \alf wave damping have been studied in
the context of spicule dynamics \citep{BDP}. In the solar
photosphere, the effect of ion-neutral damping on the propagation of
waves has also been recently studied \citep{KR}.  Our aim therefore
is to develop an approximate single fluid like description of a
multi-component, partially ionized plasma and demand that it reduces
to the fully and weakly-ionized descriptions in different fractional
ionization limits.  This approximate formulation will permit us to
explore the relationship between the onset of the Hall effect due to
the ion inertia or due to the ion-neutral collisions.  Furthermore,
such a description will provide us the freedom to investigate the
effect of fractional ionization in various limits on the MHD wave
modes.

\subsection{A single-fluid model for partially-ionized plasma}
We start with the three-component (ions, electrons and neutrals)
description of a partially ionized plasma and reduce it to a single
fluid description.  The continuity equation is
\begin{equation}
\frac{\partial \rho_j}{\partial t} + \grad\cdot\left(\rho_j\,\vj\right) = 0\,,
\label{contj}
\end{equation}
where $\rho_j = m_j\,n_j$ is the mass density, $\vj$ is the velocity,
and $n_j$ and $m_j$ are the number density and particle mass of the
various components for $j=i,\,e,\,n$.  We shall assume that the ions
are singly charged and adopt charge neutrality, so that $n_i = n_e$.
The momentum equations for the electrons, ions and neutrals are
\begin{equation}
\frac{d\ve}{dt}= - \frac{\nabla P_e}{\rho_e} - \frac{e}{m_e}\left(\E + \frac{\ve}{c}\cross \B\right)
-\!\! \sum_{j=i,n}\!\!\nu_{ej}\left(\ve - \vj \right)
\label{eeq}
\end{equation}
\begin{equation}
\frac{d\vi}{dt}=   -\frac{\nabla\,P_i}{\rho_i}  + \frac{e}{m_i}\,\left(\E + \frac{\vi}{c}\cross \B\right)
-\!\! \sum_{j=e,n}\!\!\nu_{ij}\left(\vi - \vj \right)
\label{ieq}
\end{equation}
\begin{equation}
\frac{d\vn}{dt}= - \frac{\nabla\,P_n}{\rho_n} +
\sum_{j=e,i} \!\!\nu_{nj}\left(\vj - \vn \right)\,.
\label{neq}
\end{equation}
The electron and ion momentum equations (\ref{eeq})-(\ref{ieq})
contain on the right hand side pressure gradient, Lorentz force and
collisonal momentum exchange terms where $P_j$ is the pressure, $\E$
and $\B$ are the electric and magnetic field, $c$ is the speed of
light, and $\nu_{ij}$ is the collision frequency for species $i$ with
species $j$.  The electron-ion collision frequency $\nu_{ei}$ can be
expressed in terms of the fractional ionization $x_e = n_e/n_n$  and the
plasma temperature $T_e = T_i = T$ as
\begin{equation}
\nu_{ei} = 51\,x_e\,n_n\,T^{-1.5}\,\mbox{s}^{-1}\,,
\end{equation}
where $\mbox{T}$ and $n_n$ are in $\mbox{K}$ and $\mbox{cm}^{-3}$ respectively. 
The plasma-neutral collision frequency $\nu_{j n}$ is
\begin{equation}
\nu_{j n} = \gamma_{j n}\,\rho_n  = \frac{\sigv_j}{m_n + m_j}\,\rho_n \,.
\end{equation}
Here $\sigv_j$ is the rate coefficient for the momentum transfer by
collision of the $j^{\mbox{th}}$ particle with the neutrals.  The
ion-neutral and electron-neutral rate coefficients are \citep{d6}
\begin{eqnarray}
<\sigma\,v>_{in} &=& 1.9 \cross 10^{-9}\quad \mbox{cm}^3\,\mbox{s}^{-1}\nonumber \\
<\sigma\,v>_{en} &=& 8.28 \cross 10^{-10}\,T^{\frac{1}{2}}\quad \mbox{cm}^3\,\mbox{s}^{-1}\,.
\end{eqnarray}

The density of the bulk fluid is
\begin{equation}
\rho = \rho_e + \rho_i + \rho_n \approx \rho_i + \rho_n
\label{rho}\,.
\end{equation}
Then defining the neutral density fraction
\begin{equation}
    D = \frac{\rho_n}{\rho} \,,
    \label{eq:D}
\end{equation}
the bulk velocity $ \v = (\rho_i\,\vi + \rho_n\,\vn)/\rho$ can be written as
\begin{equation}
\v = (1-D)\,\vi + D\,\vn.
\label{vt}
\end{equation}
Note that we are implicitly neglecting the electron inertia in
(\ref{vt}), and therefore in the momentum equation (\ref{eq:momentum}) below.

The continuity equation for the bulk fluid is obtained by summing up equation
(\ref{contj}) for each species:
\begin{equation}
\frac{\partial \rho}{\partial t} + \grad\cdot\left(\rho\,\v\right) = 0\,.
\label{eq:continuity}
\end{equation}
The momentum equation can be derived by adding equations (\ref{eeq})
-- (\ref{neq}) to obtain
\begin{equation}
\rho\,\frac{d\v}{dt} + \grad\cdot \left( \frac{\rho_i\rho_n}{\rho}\,\vD\vD \right)
= - \nabla\,P + \frac{\J\cross\B}{c} \,,
\label{eq:meq1}
\end{equation}
where $P = P_e + P_i + P_n$ is the total pressure, $\vD = \vi - \vn$ is the
ion-neutral drift velocity, and $\J = n_e\,e\,\left(\vi - \ve\right)$ is
the current density.

Defining $v_A = B / \sqrt{4\,\pi\,\rho}$ as the \alf speed in the bulk fluid and $c_s = \sqrt{\gamma\,p / \rho}$ as the acoustic speed, we note that if $\rho_i\rho_n v_D^2 \ll \rho^2 (v_A^2 + c_s^2)$ then we may neglect the
$\vD\,\vD$ term in Eqn~(\ref{eq:meq1}) and recover the single-fluid momentum equation
\begin{equation}
\rho\,\frac{d\v}{dt} = - \nabla\,P + \frac{\J\cross\B}{c} \,.
\label{eq:momentum}
\end{equation}
To derive a criterion for this, we estimate $\vD$  by rewriting the ion and neutral equations of motion (\ref{ieq}) and (\ref{neq}) as
\begin{eqnarray} \left(\rho_i\,\nu_{in} + \rho_e\,\nu_{en}\right)\,\vD = - \rho_i\,\frac{d\vi}{dt} - \grad \left(P_e + P_i\right) 
    \nonumber \\
    + \frac{\J\cross\B}{c} + \frac{m_e \,\nu_{en}}{e}\J\,
\label{drvi}
\end{eqnarray}
and
\begin{equation} \left(\rho_i\,\nu_{in} + \rho_e\,\nu_{en}\right)\,\vD =
\rho_n\,\frac{d\vn}{dt}+\grad P_n + \frac{m_e \,\nu_{en}}{e}\J\,,
\label{drv}
\end{equation}
respectively.
Multiplying Eq.~(\ref{drvi}) by $\rho_n$ and Eq.~(\ref{drv}) by $\rho_i$ and then adding
\begin{eqnarray}
\left(\rho_i\,\nu_{in} + \rho_e\,\nu_{en}\right)\vD = 
D\, \frac{\J\cross\B}{c} + \grad P_n ö- D\,\grad P 
\nonumber\\
+ \frac{\rho_i\,\rho_n}{\rho}\left[\frac{d\vD}{dt} - \left(\vD\cdot\grad\right)\vi ö-  \left(\vi\cdot\grad\right)\vD \right]
+ \frac{m_e\,\nu_{en}}{e}\J 
\,.
\label{eq:drf}
\end{eqnarray}
 
The term in the square bracket can be neglected if
\bq
 \omega \lesssim \frac{\rho}{\rho_i}\, \,\nu_{ni} \,.
    \label{eq:odx}
\end{equation}
Then equation (\ref{eq:drf}) can be written as,
\begin{equation}
\vD = D\, \frac{\J\cross\B}{c\,\rho_i\,\nu_{in}} +
\frac{\grad P_n}{\rho_i\,\nu_{in}} 
 - D\frac{\grad P}{\rho_i\,\nu_{in}} 
+ \left(\frac{\beta_i}{\beta_e}\right)\,\frac{\J}{e\,n_e}\,,
\label{eq:sca}
\end{equation}
where 
\bq
\beta_j = \frac{\omega_{cj}}{\nu_{j}}\,,
\eq
is the ratio of the cyclotron frequency of the $\mbox{j}^{\mbox{th}}$ particle $\omega_{cj} = e\,B/m_j\,c$ (where
$e\,,B\,,m_j\,,c$ denots electron charge, magnetic field, mass and speed of light respectively) to the 
sum of the plasma-plasma, and plasma -- neutral,  $\nu_{jn}$collision frequencies. For electrons $\nu_{e} = \nu_{en} + \nu_{ei}$ and for ions 
$\nu_{i} = \nu_{in} + \nu_{ie}$. While writing (\ref{eq:sca}), we have 
used $\rho_e\,\nu_{en}\ll\rho_i\,\nu_{in}$. In the weakly ionized limit, when $D \rightarrow 1$, neglecting plasma pressure terms and 
assuming $\beta_e \gg 1$, Eq.~(\ref{eq:sca}) reduces to the strong coupling
approximation, i.e. $\vD \approx (\J\cross\B) / (c\,\rho_i\,\nu_{in})$\citep{shu}. 

Equation (\ref{eq:sca}) implies that for gradients with a length
scale $L$, and signal speed $s$, $v_D \sim \rho_n\,s^2\,(1 + 1/D\,\beta_e) /
\left(\rho_i\,\nu_{in}\,L\right)$. The associated dynamical frequency is $\omega
\sim s/L$, so the requirement $\rho_i\,\rho_n\,
v_D^2 \ll \rho^2\,\left(v_A^2 + c_s^2\right)$ means that the
$\vD\,\vD$ term in (\ref{eq:meq1}) can be neglected for dynamical
frequencies satisfying
\begin{equation}
    \omega \lesssim \frac{\rho}{\sqrt{\rho_i\,\rho_n}}\,\left(\frac{D\,\beta_e}{1 + D\,\beta_e}\right)\, \,\nu_{ni} \,.
    \label{eq:omega1}
\end{equation}
At higher frequencies the single-fluid approximation (\ref{eq:momentum}) breaks down. Note that this frequency constraint is much weaker in
the highly-ionized and weakly-ionized limits, for which
$\rho \approx \rho_n\,(D \rightarrow 1)$. In the appendix we show that Eq.~(\ref{eq:omega1}) is a conservative bound on the dynamical frequency. Further, we also show in the appendix that (\ref{eq:odx}) is implied by Eq.~(\ref{eq:omega1}). 

To obtain an equation for the evolution of the magnetic field, we
need to derive an expression for the electric field $\E$ in terms of
the fluid properties to insert into Faraday's law
\begin{equation}
    \delt{\B} = -c \curl\E \,.
    \label{eq:Faraday}
\end{equation}
We start with the electron momentum equation (\ref{eeq}), which in the
zero electron inertia limit yields an expression for the electric
field in the rest frame of the ions:
\begin{equation}
\E + \frac{\vi}{c}\cross \B = - \frac{\nabla\,P_e }{e\,n_e} +
\frac{\J}{\sigma} + \frac{\J\cross\B}{c\,e\,n_e}
-\frac{m_e\,\nu_{en}}{e}\,\vD\,
\label{efeq}
\end{equation}
where
\begin{equation}
  \sigma = \frac{e^2 n_e}{m_e\left(\nu_{en} + \nu_{ei}\right)}
\end{equation}
is the ohmic conductivity and $\J$ is given by Amp\'ere's law,
\begin{equation}
    \J = \frac{c}{4\pi}\curl\B\,.
    \label{eq:Ampere}
\end{equation}
It is desirable to have an expression for
electric field (\ref{efeq}) in the bulk fluid frame.  To obtain this we
use $\vi = \v + D\,\vD$, with eq. (\ref{eq:sca}) for $\vD$.
Substituting the result into (\ref{efeq}) to obtain
\begin{eqnarray}
\delt \B = \curl\left[
\left(\v\cross\B\right) 
- \frac{\J\cross\B}{e\,n_e}
+ D^2\frac{\left(\J\cross\B\right)\cross\B}{c\rho_i\,\nu_{in}}
\right. \nonumber\\
\left. -  \frac{\J}{\sigma} + \frac{D^2}{\rho_i\,\nu_{in}} \left( \frac{\rho_i}{\rho_n}\grad P_n - \grad
P_i - \grad P_e\right)\cross \B \right]. 
\label{ind}
\end{eqnarray}
where we have neglected the ``Biermann's battery'' contribution from
the $\grad P_e / en_e$ term in eq.\ (\ref{efeq}) as well as small
terms of order $D\beta_i/\beta_e$, which is $\la 10^{-3}$.  The right
hand side of this induction equation has convective, ohmic, Hall and
ambipolar diffusion terms respectively.  We note that ambipolar term in a
partially ionized plasma includes a contribution from the pressure
gradient terms as well from the magnetic stresses.

The relative importance of the various terms in the induction equation
(\ref{ind}) can be easily estimated.  The ratio of the Hall $(H)$ and
the Ohm $(O)$ terms gives $H/O \sim \beta_e$, the electron Hall
parameter.  The ratio between ambipolar $(A)$ and Hall $(H)$ terms are
$A/H \sim D^2 \,\beta_i$.  In the weak ionization ($D \rightarrow 1$)
limit, $A/H \sim \beta_i$ i.e. ion Hall parameter determines the
relative importance between the Ambipolar and the Hall terms.  In a
highly ionized plasma, $D \simeq 0$ and, the ambipolar effect becomes
inconsequential.  Unlike ambipolar diffusion, Hall diffusion does not
disappear in the high fractional ionization limit.

The ambipolar diffusion terms in (\ref{ind}) arise from
$D\,\vD\cross\B$ in the $\vi\cross\B$ term in (\ref{efeq}) since $\vi\cross\B = \v\cross\B + D\,\vD\cross\B$. The terms due to pressure gradients $\grad P \cross \B$ are negligible compared to the inductive term $\v\cross\B$ when
\begin{equation}
    \omega \lesssim \left(\frac{v_A^2}{c_s^2}\right)\, \frac{\rho^2}{\rho_i\,\rho_n}\,\nu_{ni} \,,
    \label{eq:camb}
\end{equation}
where $c_s$ is some effective sound speed.  We note that for $D\,\beta_e \sim 1$, Eq.~(\ref{eq:omega1}) guarantees (\ref{eq:camb}) when $v_A \lesssim c_s$. In the opposite limit, when $v_A > c_s$, (\ref{eq:camb}) is not implied by  (\ref{eq:omega1}). Our final induction equation without $\grad P \cross \B$ term becomes:
\begin{eqnarray}
\delt \B = \curl\left[
\left(\v\cross\B\right) - \frac{4\,\pi\,\eta}{c}\,\J - \frac{4\,\pi\,\eta_H}{c}\,\J\cross\hB
\right. \nonumber\\
\left.
+ \frac{4\,\pi\eta_A}{c}\,
\left(\J\cross\hB\right)\cross\hB
\right]\,,
\label{eq:induction}
\end{eqnarray}
where $\hB = \B /B$, and the Ohmic ($\eta$), ambipolar ($\eta_A$) and
Hall ($\eta_H$) diffusivity are 
\bq \eta =
\frac{c^2}{4\,\pi\sigma}\,\,, \eta_{A} =
\frac{D^2\,B^2}{4\,\pi\,\rho_i\,\nu_{in}} \equiv 
\frac{D\,v_A^2}{\nu_{ni}}
\,, \eta_H =
\frac{c\,B}{4\,\pi\,e\,n_e}\,.
\label{eq:diffu}
 \eq
Equation (\ref{eq:induction}) is identical to the known expression for a
weakly-ionised medium (e.g.\ K\"onigl 1989) apart from the
appearance of the factor $D^2$ in $\eta_A$, which suppresses
ambipolar diffusion if the ionisation of the plasma is significant. The dependence of ambipolar term on the $D^2$ factor was first noted by \cite{cow}.  

To summarize, the single fluid equations have been derived neglecting electron inertia in the low frequency limit given by 
(\ref{eq:omega1}). Pressure gradient terms have been neglected in the induction equation (\ref{eq:induction}), which is valid if inequality (\ref{eq:camb}) is satisfied. Then equations (\ref{eq:continuity}), (\ref{eq:momentum}), and
(\ref{eq:induction}) along with prescriptions for determining $P$ and
$n_e$ describe the dynamics of a plasma of arbitrary ionization.  For
example, when the plasma is fully ionized, (i.e. $D \rightarrow 0$),
$\v = \vi$ and (\ref{eq:continuity}), (\ref{eq:momentum}), and,
(\ref{eq:induction}) reduces to the fully ionized Hall-resistive MHD
description.  In the other extreme limit $D \rightarrow 1$, the
equations reduce to those describing weakly ionized MHD (W99, BT01).
\subsection{The Hall scale}
\newcommand\meff{m_i^\mathrm{*}}

In fully ionised plasmas the Hall effect becomes important for
frequencies in excess of the ion gyrofrequency.  In natural systems
the associated time scales are usually much shorter than those of
interest and Hall dynamics can be safely neglected.  However, in
partially ionised plasmas the Hall effect becomes important on longer
length and time scales, and in weakly ionised plasmas these may even
become comparable to the dynamical time scale of the system.

This behaviour is easily inferred from the fluid equations derived in
the previous section.  If diffusion is unimportant, the characteristic
lengthscale of a gradient in the fluid  associated with frequency
$\omega$ is $L\sim v_A/\omega$ where $v_A$ is the Alfv\'en speed in
the total fluid (not just the ionized component).  Then comparing the
magnitudes of the advective and Hall diffusion terms in the induction
equation (\ref{eq:induction}), we find that the Hall term becomes
important for frequencies in excess of the Hall frequency
\begin{equation}
    \omega_H = \frac{eB}{\meff c} = \frac{\rho_i}{\rho}\,\omega_{ci}
\equiv \frac{v_A^2}{\eta_H}\,,
    \label{eq:omegaH}
\end{equation}
where the effective ion mass is
\begin{equation}
    \meff = \rho/n_e \,.
    \label{eq:meff}
\end{equation}
The corresponding Hall length scale is
\begin{equation}
    L_H = \frac{v_A}{\omega_H} =
    \left(\frac{\rho}{\rho_i}\right)^{1/2} \delta_i = \left(\frac{\rho}{\rho_i}\right)\,
\left(\frac{v_A}{\nu_{in}}\right)\,\beta_i^{-1}
    \label{eq:LH}
\end{equation}
where $\delta_i = v_{Ai}/\omega_{ci}$ is the ion skin depth with $ v_{Ai} = B /\sqrt{4\,\pi\,\rho_i}$ as the \alf speed in the ion fluid.

The Hall effect arises because through an asymmetry in the ability of
positive and negative charge carriers to drift in response to the
instantaneous electric field.  In the fully-ionized limit, for
frequencies $\omega_{ci}\la\omega\la\omega_{ce}$ electrons are able to
attain a drift velocity in instantaneous balance between electric,
magnetic and collisonal stresses, whereas the inertia of the ions
prevents them from doing so\footnote{The effects associated with
$\omega \ga\omega_{ce}$ are absent in our estimate (\ref{eq:omegaH})
because electron inertia was explicitly neglected in our development
of the single-fluid equations.}.  In the single-fluid approximation,
the ions are tightly coupled to the neutrals by collisions so that
they are unable to drift through them but must carry them along also.
Thus they pick up the neutral inertia, gaining an effective mass
$\meff$, and are unable to fully respond to changes with frequencies
in excess of $\omega_H$ (cf.\ Pandey \& Wardle 2006a, 2006b).
Implicit in this is the requirement that collisions are able to
provide the strong coupling between ions and neutrals, as noted in
eq. (\ref{eq:omega1}).

The condition $\omega\ga\omega_H$ implies that the Hall term dominates
the inductive term in (\ref{eq:induction}), but does not guarantee
that it is the dominant diffusion mechanism.  As noted earlier the
ratio of the Hall and Ohmic diffusion terms is $\sim \beta_e$,
whereas the ratio of the ambipolar and Hall diffusion terms is $\sim
D^2 \beta_i$, so for Hall diffusion to dominate the other mechanisms,
we require
\begin{equation}
    D^2 \beta_i\ll 1 \ll \beta_e \,.
\end{equation}
Note that in the weakly-ionized limit we recover the standard
requirement $\beta_i\ll 1 \ll \beta_e$.  In the fully-ionized limit
$D^2\beta_i \rightarrow 0$ and the first inequality is guaranteed.

\subsection{Magnetovorticity}

In the Hall-dominated regime, if the
effective ion mass $\meff = \rho/n_e$ is constant (in space and time)
then the concept of flux freezing can be generalised to the freezing
of magnetovorticity
\begin{equation}
    \bm{\omega_M} = \omega_H\bm{\hat{B}} + \curl\v
    \label{eq:}
\end{equation}
into the fluid flow (see e.g., the review by Polygiannakis $\&$ Moussas 2000).  To show this we take the curl
of the momentum equation (\ref{eq:momentum}) and use
the identity $(\curl\v)\cross\v = - \half\grad v^2 +
(\v\cdot\grad)\v$ to obtain
\begin{equation}
    \delt{(\curl\v)} =  \curl(\v\cross(\curl\v)) +
    \curl\left(\frac{\J\cross\B}{\rho c}\right)\,,
    \label{eq:vorticity}
\end{equation}
where $\grad\rho\cross\grad P$ has been neglected.
In the absence
of magnetic forces, this is the equation for conservation of the
vorticity $\curl\v$.  The magnetic term is directly proportional to the Hall
term in the induction equation
\begin{equation}
    \delt{\B} =  \curl(\v\cross\B) +
    \curl\left(\frac{\J\cross\B}{en_e}\right)\,,
    \label{eq:indHall}
\end{equation}
and eliminating this term between the two yields
\begin{equation}
    \delt{\,\bm{\omega_M}} =
    \curl\left(\v\cross\bm{\omega_M}\right)
    \label{eq:magnetovorticity_freezing}
\end{equation}
which shows that the magnetovorticity is frozen into the fluid.

In the limit $|\curl\v|\ll\omega_H$, this reduces to magnetic flux
freezing, because Hall diffusion is not significant in the induction
equation.  In the opposite limit, $|\curl\v|\gg\omega_H$,
$\bm{\omega_M}$ reduces to the usual fluid vorticity $\curl\v$ because
the magnetic term is unimportant in the momentum equation.

\section{Waves in a partially ionized medium}

In this section we examine the wave modes supported by a partially
ionized plasma satisfying eqs (\ref{eq:continuity}),
(\ref{eq:momentum}) and (\ref{eq:induction}). We shall assume a homogeneous, uniform background with zero flow and
investigate the wave properties of the medium in various ionization
limits.  We assume that the medium is isothermal, i.e.\ that $P=\rho
c_s^2$ with constant sound speed $c_s$.  The linearized equations for the
perturbations $\drho$,
$\dv$, $\dB$, and $\dJ$ are
\begin{equation}
\frac{\partial \drho}{\partial t} + \grad\cdot\left(\rho\,\dv\right) = 0,
\label{cl}
\end{equation}
\begin{equation}
\rho\,\frac{d\dv}{dt}= - c_s^2\,\drho + \frac{\dJ\cross\B}{c},
\label{ml}
\end{equation}
\begin{eqnarray}
\delt \dB = \curl\left[
\left(\dv\cross\B\right) -  \frac{4\,\pi\,\eta}{c}\dJ - \frac{4\,\pi\,\eta_H}{c}\, \dJ\cross\B
\right. \nonumber\\
\left.
+ \frac{4\,\pi\,\eta_A}{c}\, \left(\dJ\cross\hB\right)\cross\hB
\right]\,,
\label{lind}
\end{eqnarray}
and
\begin{equation}
    \dJ = \frac{c}{4\pi}\curl\dB\,.
    \label{eq:dJ}
\end{equation}
Note that we do not need an explicit prescription for $\rho_i$ or $n_e$
as their perturbations do not appear in the linearized induction
equation. Assuming that the perturbations have the form $\exp(i\omega
t - i\k\cdot\x)$, and using (\ref{eq:dJ}), equations (\ref{cl}) and (\ref{ml}) become
\begin{equation}
\omega\,\drho - \rho\,\k\cdot\dv = 0\,,\nonumber\\
\end{equation}
\begin{equation}
\omega\,\dv = c_s^2\left(\frac{\k\cdot\dv}{\omega}\right)\,\k
- \frac{\left(\k\cdot\B\right)\dB}{4\pi\rho}
+ \frac{(\dB\cdot\B)\,\k}{4\pi\rho}
\label{ml1}
\end{equation}
We define $\bar{\omega}^2 = \omega^2 - k^2\,c_s^2$, dot equation
(\ref{ml1}) with $\k$, and use $\k\cdot\dB = 0$ to write
\begin{equation}
\k\cdot\dv = \frac{\omega}{4\,\pi\,\rho}\frac{k^2}{\bar{\omega}^2}\left(\B\cdot\dB\right).
\label{cnt}
\end{equation}
We see that in the incompressible limit, both $\dv$ and $\dB$
are transverse to the background magnetic field $\B$.  Making use of
equation (\ref{cnt}), equation (\ref{ml1}) can be written as
\begin{equation}
\omega\,\dv = \frac{-1}{4\,\pi\,\rho} \left[\left(\k\cdot\B\right)\dB -
\left(\frac{\omega^2}{\bar{\omega}^2}\right)\left(\dB\cdot\B\right)\,\k \right].
\end{equation}
Defining $\kh\cdot\Bh = \cs$ and $\omega_A^{2} = k^2\,v_A^2$, and
eliminating $\dv$ and $\k\cdot\dv$ from equation (\ref{lind}) we get an equation
in terms of $\dB$ only
\begin{eqnarray}
\left[\omega^2 - \left( v_A^2 + i\,\eta_A\,\omega\right)\,k^2 \,\cos^2\theta
- i\, \eta\,k^2\,\omega\right]\dB = \nonumber \\
\left[ \frac{\omega^2}{{\bar{\omega}}^2}\,\omega_A^2 + i\,\eta_A\,k^2\,\omega\, \right] \left(\dB\cdot\Bh\right)\,\left(\Bh - \kh\,\cs \right)
\nonumber \\
 - i\,\eta_H\,k^2\,\omega\,\cs\,\left(\hat{\k}\cross\dB\right)\,,
\label{ME}
\end{eqnarray}
where $\omega_A = k\,v_A$ is the \alf frequency. After some straightforward algebra, following dispersion relation can be derived from equation (\ref{ME})
\begin{eqnarray}
\left[\omega^2 - \left( v_A^2 + i\,\eta_A\,\omega\right)\,k^2 \,\cos^2\theta
- i\, \eta\,k^2\,\omega\right]
\times
\nonumber\\
\left\{\left[\omega^2 - \left( v_A^2 + i\,\eta_A\,\omega\right)\,k^2 \,\cos^2\theta
- i\, \eta\,k^2\,\omega\right]
\right.
\nonumber\\
\left.
- \,k^2 \,\sin^2\theta \,
\left[\frac{\omega^2}{{\bar{\omega}}^2}\,v_A^2  + i\,\eta_A\,\omega \right] \right\}
 - \eta_H^2\,k^4\,\omega^2\,\cos^2\theta = 0\,.
\label{DER}
\end{eqnarray}
In the following sections, we shall investigate the dispersion
relation, Eq.~(\ref{DER}) in various limits.
\subsection{No Hall limit}
In the absence of Hall (i.e. $\omega \ll \omega_H$), the last term
in dispersion relation (\ref{DER}) can be ignored.  We note that the
modes related to the $\dB \parallel \B$ and to $\dB \parallel
\hat{\k}\cross\hat{\B}$ are mixed.  For example, when the magnetic
field perturbation is parallel to the background field, i.e. $\dB
\parallel \B$, from equation (\ref{ME}), we get following dispersion
relation
\begin{eqnarray}
\omega^4 - i \left( \eta_A + \eta\right)\,k^2\,\omega^3 -
\left(c_s^2 + v_A^2\right)\,k^2\omega^2 
\nonumber \\
+ i\left(k^2\,c_s^2\right) 
\left\{
\left(
\eta_A + \eta\right)\,k^2\,\omega
- i\,k^2\,v_A^2\,\cos^2\theta \right\}= 0\,,
\label{amb1}
\end{eqnarray}
which is the second curly bracket in the dispersion relation (\ref{DER}).

 In the absence of dissipation (valid for long wavelength fluctuations),
 when $\eta$ and $\eta_A \rightarrow 0$, the roots of the dispersion relation are
\begin{equation}
\omega^2 = \frac{\left(k^2\,c_s^2 + \omega_A^2 \right)}{2}\left\{  1 \pm \left[1 -
\frac{4\,k^2\,c_s^2\,\omega_A^2 \,\cos^2\theta}{\left(k^2\,c_s^2 + \omega_A^2 \right)^2} \right]^{1/2} \right\}\,.
\label{alff}
\end{equation}
The upper and lower sign of Eq.~ (\ref{alff}) correspond to the fast and slow modes of ideal MHD.

For a cold, collisional medium, the dispersion relation (\ref{amb1})
becomes
\begin{equation}
\omega^2 ö- i\,\left(\eta_A + \eta\right)\,\omega\,k^2 - \omega_A^2 = 0.
\label{dsc1}
\end{equation}
The real and imaginary part of the root of equation (\ref{dsc1}) is
\[
Re[\omega] = \pm \omega_A\,\left[ 1 - 0.5\,\left( \frac{k\,\left(\eta_A + \eta\right)}{v_A}\right)^2 \right]^{1/2}\,,
\]
\begin{equation}
Im[\omega] = 0.5\,k^2\,\left(\eta_A + \eta\right)\,.
\label{eq:art}
\end{equation}
It is well known that the waves are damped in the weakly ionized collisional medium \citep{TM62, KP, FZS88}. The damping of the waves is not only dependent on the ion-neutral collision frequency but also on the ratio of the neutral to the bulk mass densities $D$ \citep{KR}. Eq.~(\ref{eq:art}) suggest that in the absence of Ohmic dissipation, modes with the wavelength larger than 
\begin{equation}
\lambda_{\mbox{cutoff}} = \sqrt{2}\,\pi\,\frac{D\,v_A}{\nu_{ni}}\,,
\label{eq:dwl1}
\end{equation} 
can propagate in the medium. Clearly, when $D = 1$, i.e. when the medium is weakly ionized, except for a $\sqrt{2}$ factor, this expression is same as given by \cite{KP}. Thus, $D = 1$ provides the upper bound on the wavelength of the damped mode. With the increase in the fractional ionization, the cut off wavelength decreases. 

When magnetic perturbation is along $\hat{k}\cross\hat{B} = \hat{\bf
n}\,\sin \theta$, then dotting equation (\ref{ME}) with $\hat{\bf
n}\,\sin \theta$, we get following dispersion relation
\begin{equation}
\omega^2 ö- i\,\left(\eta_A\,\cos^2\theta + \eta\right)\,\omega\,k^2 - \omega_A^2\,\cos^2\theta = 0.
\label{dsc}
\end{equation}
which corresponds to the first square
bracket in equation (\ref{DER}).  The real and imaginary part of
the root of equation (\ref{dsc1}) is
\[
Re[\omega] = \pm \omega_A\,\left[ 1 - 0.5\,\left( \frac{k\,\left(\eta_A\,\cos^2\theta + \eta\right)}{v_A\,\cos\theta}\right)^2 \right]^{1/2}\,,
\]
\begin{equation}
Im[\omega] = 0.5\,k^2\,\left(\eta_A\,\cos^2\theta + \eta\right)\,.
\label{dsc2}
\end{equation}
The normal mode behaviour of the waves are similar in both
$\dB\cdot\B$ and $\dB\cdot \kh\cross\Bh$ cases except for the
$\cos\theta$ reduction factor in the later case.  We note that
equation (\ref{dsc2}) is identical to equation (17) of \cite{de}.
The damping of the magnetic fluctuations along $\hat{\bf
n}$ is reduced by the $\cs$ factor in the transverse direction.

In the limit $c_s^2 \rightarrow \infty$, we obtain the dispersion relation found by 
\cite{de}) in the Boussinesq approximation ($\delta \rho = 0\,, \delta P \neq 0$).
\subsection{Hall limit }
This dispersion relation (\ref{DER}), acquires a familiar form (cf.\ Wardle \& Ng 1999, Eq.~25) when wave is propagating along the ambient magnetic field ($\theta = 0$)
\begin{equation}
\omega^2 ö- i\,\eta_T\,k^2\omega
- \omega_A^2 = \pm \eta_H\,k^2\,\omega\,,
\label{eq:WN}
\end{equation}
where $\eta_T = \eta_A + \eta$. 
In the low frequency limit $\omega \ll \omega_A$, neglecting $\omega^2$ in 
(\ref{eq:WN}), we get
\begin{eqnarray}
Re[\omega] = \pm \omega_H \frac{1}{ 1 + \left(\beta_e^{-1} +  D^2\,\beta_i\right)^2}\,,\nonumber\\    
Im[\omega] = \omega_H\,\frac{\beta_e^{-1} + D^2\,\beta_i}{ 1 + \left(\beta_e^{-1} +  D^2\,\beta_i\right)^2}\,.
\label{eq:ICM}
\end{eqnarray}

The modified ion-cyclotron mode, Eq.~(\ref{eq:ICM}) has very low threshold of excitation in a weakly ionized medium since $Re[\omega] \approx \omega_H \approx 0$. The ratio of imaginary and real part of the frequency only in the presence of ambipolar diffusion ($\eta = 0$) gives
\begin{equation}
\frac{Im[\omega]}{Re[\omega]} = D^2\,\beta_i\,
\end{equation}  
which is same as the ratio of ambipolar to Hall term in the induction Eq.~(\ref{eq:induction}). Above expression can also be written in terms of Hall and Pedersen conductivities \citep{w5}. We note that when $\beta_i \ll 1$, damping of the waves will be insignificant and system can support very low frequency ion-cyclotron modes. 

In the presence of Ohmic diffusion only, above ratio is
\begin{equation}
\frac{Im[\omega]}{Re[\omega]} = \frac{1}{\beta_e}\,.
\end{equation}  
 Recall that $H/O = \beta_e$ in the induction equation Eq.~(\ref{eq:induction}). Therefore when $\beta_e \gg 1$, i.e. when Hall dominates Ohm, the damping of the ion-cyclotron mode is insignificant. We may conclude that in the Hall regime, i.e. when $D^2\,\beta_i \ll 1 \ll \beta_e$, weakly ionized plasma can easily excite very low frequency ion-cyclotron mode which will propagate undamped in the medium. 

Since the excitation threshold of modified ion-cyclotron wave is close to zero this mode will always be present in the medium, except when the direction of wave propagation is almost transverse to the ambient magnetic field, since for oblique propagation  $Re[\omega] \approx \omega_H\,\cos\theta$. We note that the excitation of the very low frequency modified ion cyclotron mode in a weakly ionized medium is a novel feature of the Hall MHD. This
feature makes it different from a highly ionized case.  Therefore, in the weakly ionized medium such as dark clouds and protoplanetary discs, where $\omega_H \approx 0$, the ion-cyclotron mode is likely to exist in the medium.

In the high frequency limit $\omega_A \ll \omega$, neglecting $\omega^2$ in 
(\ref{eq:WN}), and assuming $\eta = 0$, we get
\begin{eqnarray}
Re[\omega] = \pm \frac{\omega_A^2}{\omega_H}\,,\nonumber\\    
Im[\omega] = D\frac{\omega_A^2}{\nu_{ni}}\,,
\label{eq:ICM1}
\end{eqnarray}
and the ratio of imaginary and real part of the frequency gives $D^2\,\beta_i$ implying that, the system excites low frequency ion-cyclotron and high frequency whistler waves in the system when $D^2\,\beta_i \ll 1$.

We note that the nature of the whistler wave in a partially
ionized medium is different from the whistler in a fully
ionized medium. In $\omega_H \rightarrow 0$ limit, the whistler frequency can become very high but the present single fluid description is valid only for whistler frequencies satisfying equation (\ref{eq:omega1}). In terms of wavelength this constraint becomes
\begin{equation}
\lambda \gtrsim 2\,\pi\,
\left(\frac{\rho_i}{\rho_n}\right)^{1/4}\,
 \left(\frac{1 + D\,\beta_e}{D\,\beta_e}\right)^{1/2}\,
\left(\frac{\eta_H\,\cs}{\nu_{ni}}\right)^{1/2}\,.
\label{ub_whis1}
\end{equation}
We note that the above expression provides a lower bound on the wavelength. In a medium such as molecular clouds, taking $D = 1$ and
calculating $\eta_H$ for a mGauss field with $n_e \sim
.01\,\mbox{cm}^{-3},\,m_i = 30\, m_p,\,m_n = 2.35\,m_p$  
and $\nu_{ni} = 2.8 \times 10^{-12}\,\mbox{s}^{-1}$ for
$n_n \sim 10^{6}\,\mbox{cm}^{-3}$, we get
$\lambda \gtrsim 10^{4}\,\mbox{cm}$. This suggests that single fluid description permits the excitation of very small wavelength fluctuations in the cloud. 
We may conclude that the weakly ionized interstellar medium is capable of exciting high frequency, short ($k \rightarrow \infty$) whistlers in the medium.

To summarize, both the modified ion-cyclotron and whistler waves correspond to the short wavelength limit of Eq.~( \ref{eq:WN}
), i.e. $\omega_H \ll \omega_A$. In the long wavelength ($\omega_A \ll \omega_H$) limit, we get familiar Alfven wave $\omega^2 \simeq \omega_A^2$.
\section{Applications}
In this section we consider the relevance of the Hall effect in fusion, space, and astrophysical plasmas.  We do this by adopting typical parameters for the plasmas and examining the Hall length and time scales, and the relative magnitudes of Hall, ambipolar and Ohmic diffusivities.  We also discuss its likely implications.

\subsection{Fusion plasmas}
It is well known that in the fusion devices Hall can play an important role in discharge behaviour such as the sawtooth collapse of the tokamak discharge \citep{WB93}, or, in non-Ohmic current drive schemes \citep{pan}. The Hall effect is also potentially important near the partially ionized wall region of a tokamak. Since $n_n/n_i \sim \left( 10^{-3} ö- 10^{-4}\right)$ near the wall region \citep{FCH01}, it is clear from Eq.~(\ref{eq:omegaH})-(\ref{eq:LH}) that $\omega_H \approx \omega_{ci}$ and $L_H \approx \delta_i$, where we have assumed $m_n = m_i$. Clearly, Hall scaling is similar in both fully ionized core and partially ionized wall region of the tokamak. Therefore, Hall effect near the wall region will be important for $\omega_{ci} \lesssim \omega$ and scale comparable to the ion-inertial scale. 

For typical ion densities $\sim 10^{14}\,\mbox{cm}^{-3}$ and $10\, \mbox{kG}$ field in fusion plasmas, assuming $m_i = m_p$ we get $\omega_{ci} = 10^{8}\,\mbox{s}^{-2}$. The ion \alf speed is $v_{Ai} = 2.23 \times 10^{8}\,\mbox{cm}$, and, the ion skin depth is $\delta_i = v_{Ai} / \omega_{ci} \sim 2.23$ cm.  Adopting $\sim 10^{2}\, \mbox{cm}$ as the major radius of the tokamak plasma, the \alf frequency is $\omega_A \equiv R^{-1}\,V_{A} \sim 10^{7} \mbox{s}^{-1}$.  Therefore, the Hall scale $\sim \mbox{few}\,\mbox{cm}$.

\subsection{Ionospheric plasmas}
An important question for magnetosphere-ionosphere coupling is the interaction between the collisionless magnetospheric plasma and the collisional ionospheric plasma. The magnetosphere is well described by the ideal or Hall MHD equations whereas the ionosphere is described by the fluid equations along with the inertialess plasma determining the relationship between the current and the electric field through a generalized Ohm's law. The transition between the two regions is not easily facilitated using these different approaches. In particular they mask why Hall operates at large scales in the ionosphere, and shrinks to the ion-inertial scale in the magnetosphere. The unified set of equations presented here treats the ionosphere and magnetosphere in the same framework and describes the dynamics of the transition region in a consistent fashion. The added bonus of this approach is that it explains why the Hall scale 
shrinks as one moves from ionosphere to the magnetosphere.  

To illustrate these points and to gauge the relative importance of ambipolar, Hall, and Ohmic diffusion in the lower ionosphere, we present representative neutral mass density, collision frequencies \citep{chap, song}, and the corresponding Hall-beta parameters, the Hall frequency and Hall length scale in Table \ref{tab:table1}.  Molecular nitrogen and oxygen are the dominant components of the lower atmosphere, thus we have adopted a mean neutral mass $m_n = 16\,m_p$.
\begin{table*}
 \centering
 \begin{minipage}{140mm}
  \caption{\label{tab:table1} Mass Density $\rho$, the ratio of ion to neutral mass density $\rho_i / \rho$, ion-neutral, $\nu_{in}$ and electron-neutral, $\nu_{en}$ collision frequencies, ratio of the ambipolar to Hall, $\eta_A/\eta_H = D^2\,\beta_i$ and Hall to Ohm, $\eta_H/\eta_O = \beta_e$ diffusivities along with the Hall frequency $\omega_H$ and Hall scale length $L_H$ is shown in the table for different heights pertaining to Earth's lower ionosphere. A 0.3\,G magnetic field has been assumed.}
  \begin{tabular}{@{}llrrrrlrlr@{}}
  \toprule[0.12em]
h\,(km) & $\rho\,(\mbox{g}\,\mbox{cm}^{-3})$  & $\rho_i/\rho_n $ & $\nu_{in}\,(\mbox{Hz})$ & $\nu_{en}\,(\mbox{Hz})$
& $D^2\,\beta_i$ & $\beta_e$ & $\omega_H\,(\mbox{Hz})$ & $L_H\,(\mbox{km})$\\
\midrule[0.12em]
 $80$ & $10^{-8}$ & $10^{-12}$ & $7\cdot 10^{5}$ & $10^{7}$ & $10^{-3}$ & $.07$ &  $10^{-10}$ & $10^{7}$ &  \\
 $100$ & $10^{-9}$ & $10^{-10}$ & $7\cdot 10^{3}$ & $10^{5}$ & $0.7$ & $70$ & $10^{-8}$ & $10^{5}$ &  \\
 $130$ & $10^{-10}$ & $10^{-7}$ & $10^{2}$ & $3\cdot 10^{3}$ & $1$ & $10^{3}$ & $10^{-5}$ & $10^{4}$ &  \\
 $150$ & $10^{-11}$ & $10^{-4}$ & $30$ & $8\cdot 10^{2}$ & $3$ & $10^{4}$ & $10^{-2}$ & $10$ &  \\
\bottomrule[0.12em]
\end{tabular}
\end{minipage}
\end{table*}
Hall and Ohmic diffusion are dominant in the lower E-layer of the
ionosphere.  Above $\sim 100\,$km, Hall diffusion is dominant $\omega_H$ is very low and the corresponding Hall scale is
very large (Table \ref{tab:table1}). With increasing height, the density ratio $\rho_i/\rho$ increases and Hall length shrinks becoming of the order of ion-inertial scale when $\rho_i /\rho \sim 1$ in the magnetosphere. 

We see from Table \ref{tab:table1} that since ratio of the ambipolar (A) and Hall (H) terms are, $A/ H \equiv \eta_A/\eta_H = D^2\,\beta_i \sim 1$, both these diffusion will operate on an equal footing towards the upper E-layer ($\sim 130\,\mbox{km}$) and lower F-layer ($\gtrsim 150\,\mbox{km}$), whereas Ohmic diffusion will be unimportant. Observations of the partially ionized D and E regions close to the
lower boundary of the Earth's ionosphere ($\sim70 - 140$\,km), reveal
the permanent presence of ULF waves.  In the E-region of the
ionosphere, these waves have slow and fast components with phase
velocities between $1-100\,\mbox{m\,s}^{-1}$ and $2 -
20\,\mbox{km\,s}^{-1}$ and frequencies between $10^{-1} -
10^{-4}$\,Hz and $10^{-4} - 10^{-6}$\,Hz respectively, with
wavelength $\ga 10^{3}$\,km and a period of variation ranging between
few days to tens of days \citep{zhou, bauer}.  Day and night time
observations gives an order of magnitude difference in the phase
velocity.

The slow ULF waves have been identified as \alf waves, which due to presence of neutrals, converts to whistler waves in the E-layer \citep{abu}.   
We note that in the lower layer of the ionosphere (D, E
and lower F -layers, $\sim 70 - 140 \,\mbox{km}$), the ionization mass
fraction could be as low as $\sim 10^{-12}$ (Table \ref{tab:table1}).
Hall MHD is applicable in this region, and for $B = 0.3$\,G and an ion
mass $m_i \equiv m_{O^+} \sim 10^{-23}\,\mbox{g}$, the Hall criterion
is $\omega \ga \omega_{H}=10^{-9}$\,Hz rather than the requirement
$\omega\ga\omega_{ci}\sim 10^{3}\,\mbox{s}^{-1}$ that would hold if
the medium were fully ionized. Thus, waves in the $10^{-1} -
10^{-4}$\,Hz range are most likely whistler waves. Identifying the observed wave speed $\left(1-2\,\mbox{km}\right)$ \citep{abu} with the whistler, we obtain the characteristic wavelength $\lambda \sim 10^{3}\,\mbox{km}$ at $130\,\mbox{km}$.          
Therefore, what is being identified as slow \alf mode converting to whistler could be just low frequency whistler mode without any mode conversion.

In fully-ionized plasmas the Hall effect considerably modifies the
classical Kelvin-Helmholtz (KH) and Rayleigh-Taylor (RT) instabilities
on the ion-inertial scale \citep{TK65}. Both these instabilities grow faster in the presence of Hall effect. We anticipate that similar
modifications to the KH and RT instabilities will be effective on much
longer scales and that the Hall effect may facilitate the energy
cascade from large to small scales in the weakly-ionized regions of
the ionosphere.

\subsection{Solar atmosphere}
The potential role of the Hall effect in the solar atmosphere has escaped the attention of solar community owing to the confusion about the Hall scaling in  partially ionized plasmas. The interaction between the partially ionised solar atmosphere and fully ionized corona has important consequences for the wave heating of the corona. MHD waves can be easily excited in the medium by e.g. convective gas motions. A tiny fraction of this energy carried by the waves to higher altitude would suffice to heat the corona to high temperatures \citep{priest}. However, the dynamics of the weakly ionized photosphere is dominated by collisional effects whereas highly ionized corona is described by ideal MHD. It is unclear how does one makes a transition between the collisional photosphere and collisionless corona. The present formulation not only allows us to investigate the dynamics of the collisional lower photosphere but also allows taking the proper limit to the fully ionized coronal region. As we shall see below, Hall may indeed become important in the solar photosphere. To show this, we give typical collision frequencies and Hall parameters for the solar
atmosphere in Table \ref{tab:table2} for standard solar photosphere
and chromosphere models \citep{VAL81, cox}. The proton-hydrogen $\left(H^+ - H \right)$ elastic collision cross section is temperature-dependent and at $0.5\,eV$ is typically $2\,\cdot\,10^{-14}\,\mbox{cm}^2$ \citep{KS99}. The 
electron-hydrogen $\left(e^- - H \right)$ collision cross-section is  also temperature-dependent and is $3.5\,\cdot\,10^{-15}\,\mbox{cm}^2$ at $0.5\,eV$ \citep{BK71, ZKB96}. Table \ref{tab:table2} somewhat underestimates the collision frequencies as charge exchange between ionized and neutral hydrogen in the solar atmosphere has a large collision cross section $\sim 5.6\,\cdot\,10^{-15}\,\mbox{cm}^2$ 
\citep{KS99}.  To calculate the Hall parameters, we have assumed a magnetic field $B = 100$\,G. We note here that in the solar photosphere electron ö- ion collisions are as significant as plasma ö- neutral collisions. This could be accounted for by multiplying the plasma ö- neutral collision frequencies in  Table \ref{tab:table2} by two. However, this does not change the overall conclusions of this work.

Ohmic diffusion dominates in the quiet-Sun photosphere.  However,
given the uncertainty about the collision frequencies and
strength of the magnetic field, Ohmic and Hall diffusion may be comparable at the base of the chromosphere.
Then the Hall effect may significantly affect the excitation and
propagation of waves in the photosphere and chromosphere. As is seen from Table
\ref{tab:table2}, near the surface of the Sun, the Hall effect will
operate over few meters whereas near the base of the chromospheres,
the typical Hall scale may be of the order of few hundred kilometres. We note that for stronger magnetic fields the Hall scale
will be considerably modified.  For example, for a $1$\,kG field,
the Hall scale ranges from tens of km near the surface of the Sun to
hundreds of km as approaching the chromosphere.

\begin{table*}
 \centering
 \begin{minipage}{140mm}
  \caption{\label{tab:table2}Same as in Table \ref{tab:table1}, but
  for the solar atmosphere. A 100\,G magnetic field has been assumed.}
  \begin{tabular}{@{}llrrrrlrlr@{}}
\toprule[0.12em]
h\,(km) & $\rho\,(\mbox{g}\,\mbox{cm}^{-3})$  & $\rho_i/\rho_n $ & $\nu_{in}\,(\mbox{Hz})$ & $\nu_{en}\,(\mbox{Hz})$
& $ D^2\,\beta_i $ & $\beta_e$ & $\omega_H\,(\mbox{Hz})$ & $L_H\,(\mbox{km})$\\
\midrule[0.12em]
 $0$ & $2.77\cdot10^{-7}$ & $10^{-4}$ & $1.6\cdot 10^{9}$ & $1.3\cdot 10^{10}$ & $10^{-4}$ & $10^{-1}$ & $10$ & $10^{-1}$ &  \\
$525$ & $4.87\cdot10^{-9}$ & $10^{-4}$ & $2.2\cdot 10^{7}$ & $2\cdot 10^{8}$ & $10^{-2}$ & $1$ & $10$ & $10$ &  \\
$1000$ & $5.07\cdot 10^{-11}$ & $10^{-3}$ & $2.2\cdot 10^{5}$ & $2\cdot 10^{6}$
& $1$ & $30$ & $10^{2}$ &  $10$ &  \\
\bottomrule[0.12em]
\end{tabular}
\end{minipage}
\end{table*}

Heating of the solar corona by MHD surface waves is a plausible
mechanism for explaining the high coronal temperature
\citep{priest}. Waves in the corona are thought to have emerged from the lower photosphere, where they may have been excited by footpoint motion of the
magnetic field lines.  \alf waves propagate to the corona and lose
energy by resonance damping, and heat the plasma.  However, the lower solar
photosphere is comparatively cold ($T \sim 6000\,K$) and weakly
ionized, and ion-neutral collisional dynamics may play an important
role. We see from the values of $\beta_j$ in Table \ref{tab:table2} 
that except at $h = 0$ where Ohm probably competes with Hall, Hall is the dominant diffusion mechanism in the medium. For a field $ > 10^2\,\mbox{G}$  (a field probably present in the network, internetwork, plague and sunspot \cite{BTS01, DKS03}) even at $h = 0$ Hall will be the dominant mechanism. Ambipolar diffusion is unimportant in the lower photosphere 
($ \leq 1000\,\mbox{km}$), where $\beta_i \ll 1$. At higher altitudes, typically between 
$\left(1000 - 3000\right) \,\mbox{km}$, ambipolar diffusion will be important, above this the role of ambipolar diffusion diminishes as the neutral number density plummets rapidly. We infer that 
Hall effect is present over the entire solar atmosphere although with shrinking spatial scales with increasing altitude and will modify the  large scale wave motion in the medium.

Granulations or convective motions are believed to generate \alf waves in the photosphere.  The \alf wave is a promising candidate for heating and acceleration of the solar plasma from coronal holes. High-frequency ($\sim 10^4   \mbox{Hz}$) ion-cyclotron waves have been proposed as a candidate for preferential 
heating of the heavy ions \citep{K98}. The power spectra of horizontal photospheric motions suggest
that waves with frequencies ($10^{-5}- 0.1$\,Hz) are present at a few
solar radii \citep{cb}. Thus, it should be possible to observe
$1$\,Hz waves.  Whistlers may be excited close to the footpoint of the
flux tube, and with decreasing neutral density, this wave will turn
into an \alf wave that can propagate to the corona and heat it. As the damping of the mode decreases with the increasing ionization, the
waves can propagate almost undamped right up to the corona.  Thus, in
the lower part of the solar corona, Hall effect may generate whistler waves.  The Hall effect operates on
scales $L_H$ in excess of a few tens of kms, the wave modes due to Hall will be important to various coronal heating models. In the fully-ionized solar wind, the role of Hall term is constrained
by the smallness of the ion skin depth although even then the Hall
effect appears to significantly modify the surface wave properties in
such a plasma \citep{zhel}.

\subsection{Protoplanetary Discs}
The role of Hall drift on the dynamics of protoplanetary discs (PPDs) has been investigated in several papers (W99, BT01, \cite{ss1, ss2, sw1, sw2, pwa}). 
Table \ref{tab:table3} gives the collision frequencies, plasma Hall
parameters and Hall length and time scales in a protostellar disc, for
a minimum-solar-mass nebula model at 5 AU from the central star in the
presence of a $10^{-2}\,G$ field, using the ionization fraction
calculated by \cite{wp}.
\begin{table*}
 \centering
 \begin{minipage}{140mm}
  \caption{\label{tab:table3}Same as in table\ref{tab:table1}, but
  for 1\,AU in a protoplanetary disc. We assume $m_i = 30\,m_p$, $m_n = 2.3\,m_p$ and $B = 10^{-2} G$ corresponding to 
$\omega_{ci} \sim 1\,s^{-1}$ and $\omega_{ce} \sim 10^{5}\,s^{-1}$. $z/h$ is the no.\ of scale heights above the disc midplane.}
  \begin{tabular}{@{}llrrrrlrlr@{}}
  \toprule[0.12em]
z/h & $\rho$\,(g\,cm$^{-3}$)  & $\rho_i/\rho_n$ & $\nu_{in}$(Hz) & $\nu_{en}$(Hz)
& $\eta_A/\eta_H = D^2\,\beta_i$ & $\eta_H/\eta = \beta_e$ & $\omega_H\,$(s$^{-1}$) & $L_H\,$(km)\\
\midrule[0.12em]
 $0$ & $10^{-8}$ & $10^{-10}$ & $3\cdot 10^{3}$ & $3\cdot 10^{3}$ & $10^{-3}$ & $10^{2}$ & $10^{-10}$ & $10^6$ &  \\
$1$ & $10^{-9}$ & $10^{-7}$ & $10^{3}$ & $2\cdot 10^{3}$ & $10^{-3}$ & $10^{2}$ & $10^{-7}$ & $10^{4}$ &  \\
$3$ & $10^{-11}$ & $10^{-6}$ & $10$ & $15$& $1$ & $10^{3}$ & $10^{-6}$ &  $10^{4}$ &  \\
\bottomrule[0.12em]
\end{tabular}
\end{minipage}
\end{table*}
For these parameters, Hall diffusion is important for $\omega \ga
\omega_{H} \sim 10^{-6} - 10^{-8} \,\mbox{s}^{-1}$ comparable to the orbital frequency.  
Near the mid-plane of the PPDs Hall diffusion will be important
whereas towards the surface of the disc, ambipolar diffusion becomes dominant.
The Hall scale $L_H \sim 10^{5}\,\mbox{km}$ is comparable to tye disc thickness, uggesting that Hall operates over the large part 
of the disc.

\section{Summary}
Weakly and fully ionized plasmas often exist side by side in nature.
The transition from weakly to fully ionized plasma is typically not
abrupt but occurs smoothly, and the transition from one to the other
has posed a considerable theoretical difficulty, requiring separate
treatments of Ohm's law for weakly (with zero plasma inertia) and
fully ionized plasmas.  In this paper we have developed a
consistent MHD framework for fully, partially, and weakly ionized plasmas.
Our main results are as follows:
\begin{enumerate}
	\item  Our single fluid description encompasses the continuity, momentum and
induction equations (\ref{eq:continuity}), (\ref{eq:momentum}), and
(\ref{eq:induction}), along with simple expressions for the Ohmic, Hall and
ambipolar diffusivities specified by eqs.~(\ref{eq:diffu}).
These, along with prescriptions for determining $P$ and
$n_e$, specify the dynamics of a partially ionized plasma.
The equations neglect electron inertia
limit, and are restricted to low frequencies, i.e. $\omega \lesssim
\sqrt{\rho_n/\rho_i}\,\beta_e\,\nu_{ni}/\left(1 + D\,\beta_e\right)$.

	\item  In a partially ionized plasma the frequency $\omega_H$ above which 
the Hall effect becomes important and the corresponding spatial scale
$L_H$ depend on the
fractional ionization, i.e.
\begin{eqnarray}
\omega_H \sim \frac{\rho_i}{\rho}\,\omega_{ci}\,\nonumber\\
L_H \sim \sqrt{\frac{\rho}{\rho_i}}\,\delta_i \equiv \frac{v_A}{\omega_H}\,.
\label{eq:HsX}
\end{eqnarray}
This occurs because the ions are effectively coupled to the neutrals
by collisions, so that the effective ion mass is
$\left(\rho/\rho_i\right)\,m_i$.  This behaviour explains why the Hall
scale is negligibly small in fully ionized plasmas and yet is large in
protoplanetary discs, and the Earth's ionosphere.  It also predicts
that the Hall effect plays an important role in the dynamics of the solar
photosphere.

	\item We derived a general dispersion relation and showed that when
ambipolar dominates both Hall ($D^2\,\beta_i \gg 1$) and Ohmic ($D^2\,\beta_e\,\beta_i \gg 1$) diffusion, only waves with wavelength larger than a 
cutoff value (Eq.~(\ref{eq:dwl1})) can propagate in the
medium.  This cutoff wavelength depends on the ratio $\rho_n/\rho$.
Low frequency modified ion-cyclotron modes ($\omega = \omega_H$) and
high-frequency whistler ($\omega = \omega_A/\omega_H$) are the normal
modes of the medium in the Hall regime.  A weakly ionized medium, with
$\beta_i << 1$ will always support these waves owing to negligible
damping ($\sim D^2\,\beta_i << 1$).
\end{enumerate}
Finally, we emphasise that the single-fluid formulation here will
prove useful in studying the coupling between  the Earth's ionosphere and
magnetosphere and also between the different solar layers of solar atmosphere.

 \section*{Acknowledgments}
It is our pleasure to thank Steven Desch for pointing out an error in
our interpretation of the results of \cite{de} in an earlier version of
the paper.  This research has been supported by the Australian
Research Council and grants awarded by Macquarie University.

\appendix

\section[]{ }
In this section we derive a general criteria for the validity of the single fluid description, i.e. condition under which terms $\sim v_D^2$ in Eq.~(\ref{eq:meq1}) can be neglected. Further, we show that the validity condition of the single fluid description ensures the validity of equation (\ref{eq:odx}) which allows us to write the induction equation (\ref{ind}).  
We note from Eq.~(\ref{eq:drf}) 
\begin{equation}
v_D \sim \frac{\omega\,v_A}{\left(\nu_{ni} + \frac{\rho_i}{\rho}\,\omega\right)}\, \left(\frac{ 1 + D\,\beta_e}{D\,\beta_{e}}\right) \,,
\end{equation}
and the single fluid description, Eq.~(\ref{eq:momentum}) is valid provided $\rho_i\,\rho_n \,v_D^2 \ll \rho^2\, \left(v_A^2 + c_s^2\right)$, i.e. 
\bq
    \frac{\omega}{\nu_{ni}} \lesssim \frac{\rho}{\sqrt{\rho_i\,\rho_n}}\, \,\left( 1 + \frac{\rho_i}{\rho}\,\frac{\omega}{\nu_{ni}} \right)\,
\left(\frac{D\,\beta_e}{1 + D\,\beta_{e}}\right) \,.
    \label{eq:omega2}
\eq
In the following discussion, without loss of generality, we shall assume 
$D\,\beta_e /(1 + D\,\beta_e) \sim 1$. The argument is easily generalized for arbitrary values of this ratio. It is clear that condition (\ref{eq:omega2}) is less restrictive than Eq.~(\ref{eq:omega1}). However, it is not clear if induction Eq.~(\ref{ind}) which has been derived 
by assuming an expression for $\vD$, Eq.~(\ref{eq:sca}), under condition (\ref{eq:odx}) is consistent with Eq.~(\ref{eq:omega2}). In order to check the consistency of condition (\ref{eq:omega2}) with induction Eq.~(\ref{ind}), let us assume that condition (\ref{eq:odx}) is not valid, i.e. 
\bq
\frac{\omega}{\nu_{ni}} \gtrsim \frac{\rho}{\rho_i}\,.
    \label{eq:odx1}
\eq         
Writing Eq.~(\ref{efeq}) in the bulk frame,
\begin{eqnarray}
\E + D\,\frac{\vD\cross \B }{c} = - \frac{\v\cross \B }{c} - \frac{\nabla\,P_e }{e\,n_e} +
\frac{\J}{\sigma} + \frac{\J\cross\B}{c\,e\,n_e}\nonumber\\
-\frac{m_e\,\nu_{en}}{e}\,\vD\,.
\label{eq:eff}
\end{eqnarray}
We note that if $D\,\vD \gtrsim v_A$ then equation (\ref{eq:eff}) is not valid implying invalidity of equation (\ref{ind}). Condition $D\,\vD \gtrsim v_A$ 
implies
\bq
\frac{\omega}{\nu_{ni}} \gtrsim \frac{\rho_i}{\rho_n}
\,\left( \frac{\omega}{\nu_{ni}} + \frac{\rho}{\rho_i} \right)\,.
    \label{eq:nc1}
\eq
Clearly under condition (\ref{eq:nc1}), we are not allowed to write the induction equation (\ref{ind}).          
We consider two cases: (i) $\rho_i \gtrsim \rho_n$ and (ii) $\rho_i < \rho_n$ and show that condition (\ref{eq:nc1}) is incompatable with single fluid condition (\ref{eq:omega2}).  When $\rho_i \gtrsim \rho_n$, it is straightforward to see that condition (\ref{eq:nc1}) is never satisfied. Thus induction Eq.~(\ref{ind}) is valid when under a general single fluid condition, Eq.~(\ref {eq:omega2}). When  $\rho_i < \rho_n$, we note that
\bq
\frac{\omega}{\nu_{ni}} \gtrsim \frac{\rho}{\rho_i} \gtrsim  \sqrt{\frac{\rho_i}{\rho_n}}\frac{\rho}{\rho_i}\,,
    \label{eq:s1}
\eq         
and
\bq
\frac{\omega}{\nu_{ni}} \gtrsim \frac{\rho_i}{\rho_n}\frac{\omega}{\nu_{ni}}\,.
    \label{eq:s2} \eq
Adding Eqs.~(\ref{eq:s1}) and (\ref{eq:s2}), we get 
\bq
\frac{\omega}{\nu_{ni}} \gtrsim \sqrt{\frac{\rho_i}{\rho_n}}\,\left(\frac{\omega}{\nu_{ni}} + \frac{\rho}{\rho_i}\right)\,.
    \label{eq:nc2}
\eq         
Clearly, condition (\ref{eq:odx1}) with $\rho_i < \rho_n$ implies that Eq.~(\ref{eq:omega2}) is violated. Thus we have reached a contradiction. Therefore, the induction equation (\ref{ind}) is valid under a more general  condition, Eq.~(\ref{eq:omega2}).

In order to graphically sketch condition (\ref{eq:omega2}), we  write Eq.~ (\ref{eq:omega2}) as an equality by multiplying right hand side by a small factor $f$. Assuming $f = 0.1\,D\,\beta_e/\left(1 + D\,\beta_e\right)$, $\alpha = \sqrt{\rho_i/\rho_n}$ and $y = \omega/\nu_{ni}$,  Eq.~(\ref{eq:omega2}) becomes
\bq
y = \frac{ f\,\alpha\left(1 + \alpha^{-2}\right) }{1 ö- f\,\alpha}\,.
\label{eq:px1}
\eq

We can rewrite Eq.~(\ref{eq:omega2}) in the following form
\bq
    \frac{\omega}{\nu_{ni}} \lesssim 
\frac{f\, \frac{\rho}{\sqrt{\rho_i\,\rho_n}}\,\left(\frac{D\,\beta_e}{1 + D\,\beta_{e}}\right)}
{1 ö- f\,\left(\frac{D\,\beta_e}{1 + D\,\beta_{e}}\right)\,\sqrt{\frac{\rho_i}{\rho_n}}}\,.
    \label{eq:omega3}
\eq
\begin{figure}
     \includegraphics[scale=0.45]{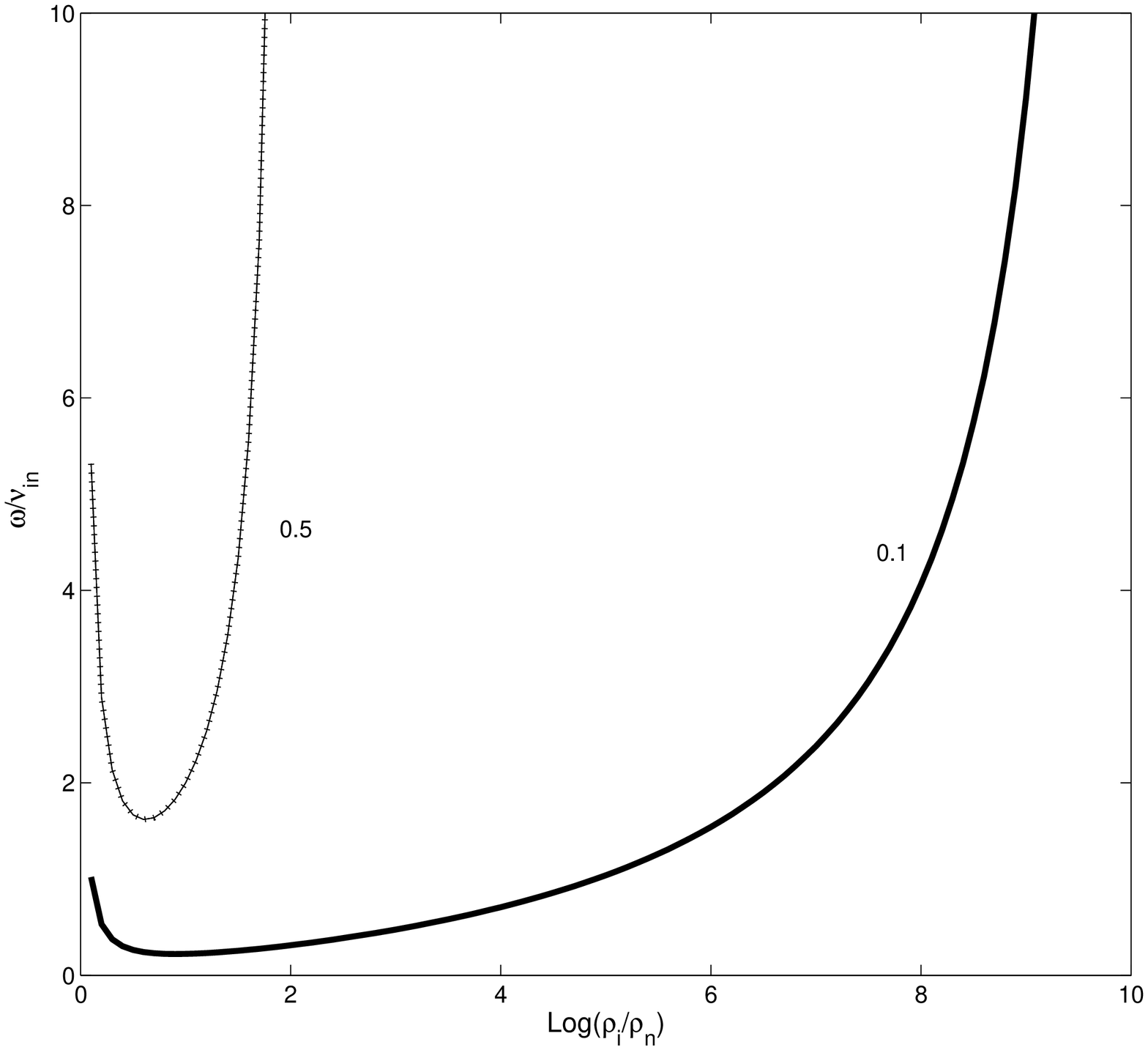}
     \caption{The numerical solution of the equation (\ref{eq:px1}) for $D\,\beta_e / \left(1 + D\,\beta_e\right) = 1\, \mbox{and}\,5$ corresponding to $f = 0.1$ and $0.5$ respectively.}
 \label{fig:freq1}  
\end{figure}
In fig.~1, we plot $\omega/\nu_{ni}$ against $\log(\rho_i/\rho_n)$ using Eq.~(\ref{eq:px1}) for $f = 0.1$ and $f = 0.5$. Both in weakly ($\rho_i \ll \rho_n $), and highly ($\rho_n \ll \rho_i$) ionized limits we see from fig.~1 that $\omega \gg \nu_{ni}$. In the weakly ionized limit, $\omega/ \nu_{ni} \sim \sqrt{\rho_n/\rho_i}$ and thus when $\rho_i/\rho_n \rightarrow 0$, $\omega$ becomes arbitrary large (fig.~1). In the highly ionized limit, $\omega/ \nu_{ni} \sim 1/\left(1 ö- f\,\rho_i/\rho_n \right)$. This results in arbitrary large $\omega$ when $\rho_n/\rho_i \rightarrow f$. Therefore, the dynamical frequency of a partially ionized plasma is bounded between the frequencies of weakly and fully ionized plasmas. 
\end{document}